# An In-router Identification Scheme for Selective Discard of Video Packets

Ashkan Moharrami, Mohammad Ghasempour and Mohammad Ghanbari, *Life Fellow, IEEE*

*Abstract*— **High quality (HQ) video services occupy large portions of the total bandwidth and are among the main causes of congestion at network bottlenecks. Since video is resilient to data loss, throwing away less important video packets can ease network congestion with minimal damage to video quality and free up bandwidth for other data flows. Frame type is one of the features that can be used to determine the importance of video packets, but this information is stored in the packet payload. Due to limited processing power of devices in high throughput/speed networks, data encryption and user credibility issues, it is costly for the network to find the frame type of each packet. Therefore, a fast and reliable standalone method to recognize video packet types at network level is desired. This paper proposes a method to model the structure of live video streams in a network node which results in determining the frame type of each packet. It enables the network nodes to mark and if need be to discard less important video packets ahead of congestion, and therefore preserve video quality and free up bandwidth for more important packet types. The method does not need to read the IP layer payload and uses only the packet header data for decisions. Experimental results indicate while dropping packets under packet type prediction degrades video quality with respect to its true type by 0.5-3 dB, it has 7-20 dB improvement over when packets are dropped randomly.**

## I. INTRODUCTION

TODAY, video is the dominant data type transported over the internet. Cisco predicts [1], by 2022 online video will comprise more than 82% of the internet traffic and live video will be about 17% of that amount. With the current pandemic and work from home policies, it is expected video over IP networks to grow even further. Also, the advent of IPTV and entertainment services through distributors like Netflix, video broadcasting over the air is diminishing and video distribution over computer networks has gained prominence. Since video volume is large by nature, the huge demand for high quality video services will put a burden over packet networks. This heavy traffic may overload the router buffers and congestion may occur during which new arriving packets of other services sharing the network will be dropped and all data flows will be adversely affected.

Several methods have been introduced in the literature to alleviate video packet loss in IP networks. Network congestion could be confronted at either the end-systems or the network itself and the approaches could be passive or active. Therefore, these methods can be divided into four major categories: passive end-system, active end-system, passive network-based, and active network-based methods. Generally speaking, passive methods do not forestall congestion and come into action only to protect data streams during congestion; whereas active methods can be used to prevent or mitigate congestion. These four categories are explained below.

*A. Passive End-system Methods*

These methods are mainly implemented in video codecs under the name of robust video coding. Examples of these methods include: layered video coding, multiple description coding (MDC) and their combinations [2], flexible macroblock ordering (FMO) [3] and interleaving [4]. The Efficiency of these methods heavily depends on the effectiveness of the used error concealment technique, where a part of the received video data, helps to reconstruct missing parts [5].

As these methods do not use feedback from the network, they can be used anywhere, especially in wireless networks. It should be noted that wireless networks are highly error prone, where in addition to data loss, erroneous packets can also be regarded as lost packets. However, if the employed entropy coding is symmetric, like the one used in H.263, the damaged area of erroneous packets can be limited to a small area, improving the efficiency of error concealment [6].

Although these methods cannot fully combat congestion in the network, nevertheless they have proven to be useful in simple environments under small packet loss rates. For instance, despite the deficiency of delay variation between base and enhancement layers in layered video coding, this coding strategy has proven to be useful and these days all the standard codecs employ such property. The implementation in both codecs and networks is simple, such that for example even older codecs, like H.261 can be easily adapted to have such property [7].

*B. Passive Network-Based Methods*

These methods are implemented at network nodes, giving routing priority to some streams over the others. The most common example of these priority-based routings is diffserv [8], [9] which can be generally categorized under QoS-based methods. Priority-based routing methods can help discriminate between different packet flows, but there are some drawbacks. For instance, in a network with competitive content distributors, the sender could not be trusted and participants may not follow a common standard for tagging their packets. Moreover, priority-based routing works solely on streams not packets.

The authors are with the School of Electrical and Computer Engineering, College of Engineering, University of Tehran, 1439957131 Iran. Emails: a.moharrami@ut.ac.ir, m.ghasempour@ut.ac.ir, ghan@ut.ac.ir

Mohammad Ghanbari is also Emeritus Professor at the School of Computer Science and Electronic Engineering, University of Essex, Colchester CO4 3SQ, UK, E-mail ghan@essex.ac.uk.

Therefore, most of the times, rather than packets, the streams are preferred over each other. Even if video packets are labeled according to their importance, since queues in network nodes are served one by one, packets will arrive with extra delay and jitter, which is very harmful for live video streams.

*C. Active End-system Methods*

The main difference between active and passive methods is that active methods simply react to certain incidents whereas passive methods are running persistently, regardless of network condition. Active end-system methods use information from the network to adapt data transmission to network condition. Examples are Dynamic Adaptive Streaming over HTTP (DASH) [10], Rate control [11]-[13], Receiver buffer analysis [14], Packet retransmission [15], Resource allocation [16] and packet discard [17]. These methods require some sort of feedback from either the network or the sending/receiving hosts and thus face possible delay issues. Many of the methods that fall under this category heavily depend on the available protocols or amendments made to the existing ones [18], [19].

*D. Active Network-Based Methods*

Since these methods are applied to network nodes, the very first points that detect congestion, they can prevent congestion by coming into action beforehand. Therefore, they are more flexible and if they are utilized properly, they can react reliably during network congestion. Examples are active queue management (AQM) [20] and selective packet discard (SPD) [21].

In active queue management methods, packets are dropped randomly according to their traffic classes whereas in selective packet discard methods, individual packets are prioritized based on a predefined function which is wholly implemented in the network node. Many different criteria could be considered to prioritize video packets over each other. Delay budget [22], frame type [23]-[27], and general contribution to video quality are some examples [28], [29].

In SPD methods, the information needed for the node to determine a packet priority is obtained from the packet itself, mainly from layers above the IP layer and then a hierarchical drop policy is implemented. For example, if the criterion is based on packet type, drop of B-type packets might ease congestion in mild conditions, but in more severe cases, P-type packets might also become a target. Loss of I-type packets is very harmful to the perceived video quality, but it too can be considered for drop in heavily congested nodes.

In methods that exercise selective packet drop, some noticeable drawbacks can be seen. For example, methods like [23] assume packet type is already known and develop their algorithms upon that. In some cases, like [27], higher layer data are considered to be available (data are not encrypted) and processing delay is negligible. Also, some may assume the encoder can mark the packet type (I & P as important) and B-type as less important packets. However, such a policy may not be fair in the current competitions among the video service providers. If, packet type is left to encoder decision, then for good delivery of video, any service provider may declare its packet type as important. This surely does not lead to optimum video delivery over the network. What is needed, packet type prediction should be left to the router. Most likely such decision, is only carried out at the network edge and packets are predicted and marked for the following routers. At the worst case, such prediction should be made at any following router.

In this paper, an in-router scheme for packet type identification and selective discard of video packets is presented to overcome the above-mentioned problems. Our method identifies packet type by reading their header data and can execute a selective packet drop based on frame type criterion in case of congestion.

Our contributions can be summarized as:
1) Presenting a fast, standalone, in-router and codec-transparent method to model the video structure and identify packet types, without inspecting the packet payload, which makes it possible to work on encrypted data flows.
2) Providing solutions for both general video packetization methods: constant coding unit per packet and constant packet size.
3) Analyzing the effect of different parameters that have impact on packets identification accuracy and video quality such as the amount of jitter, different drop rates under three different schemes: random drop, ground truth drop and the proposed smart drop methods.

The rest of the paper is organized as follows: In section II, legacy video over IP networks is reviewed. In section III, the proposed scheme is described and in section IV, experimental results are presented and finally, section V draws the concluding remarks.

## II. LEGACY VIDEO OVER IP NETWORKS

In standard video codecs such as H.264/AVC [30] and H.265/HEVC [31], a frame is made up from one or several slices that are independently decodable. Slices can be of certain number of coding units (e.g. macroblocks) or constant size in bits/bytes. In the former, due to the variations in the video content and parameters of the encoder, each slice size will be different and in the latter, each slice is comprised of several coding units such that its size remains almost constant around a certain value. However, in both cases, it is important that a coding unit not to be fragmented into two slices/packets to minimize inter-packet dependency and the side effect of lost packets [32].

In both cases, depending on the content and compression ratio, the number of bits generated per frame will be variable. This is not only due texture variation, scene content and the motion but also each frame is encoded differently. Some frames are more compressed because they use both temporal and spatial correlation, which makes the coding process more complex and as a result it takes more time to be coded. Whereas, there are frames that use only spatial correlation; hence they are less compressed and the processing time required for coding them is short. In fact, within a scene, frame to frame bit rate variation is limited and depends only on the type of the encoded frame [33], [34].

In both aforementioned codecs, network abstraction layer (NAL) is independent of video coding layer (VCL). Video transport in network is based on NAL unit which is simply a slice accompanied with a header. Usually, each NALU is encapsulated into one RTP packet and RTP packets are transmitted over UDP/IP [35]-[37]. IP packets priority can be specified in their header, e.g. in IPv4 a 3-bit precedence field located in the type of service (ToS) byte indicates the packet priority and in IPv6 the traffic class can be reflected in the differentiated services code point (DSCP) field. An IPv6 packet that contains a NAL unit is shown in Figure 1.

Video parameters are hidden inside the NALU and are not reflected in lower layers. Therefore, network nodes cannot know video packet type, unless they read the higher layers header and payload content, which is time consuming and requires heavy processing specially for encrypted payloads. They may only rely on traffic class at packet header, but this cannot be trusted in a competitive world of service providers sharing the network.

### III. THE PROPOSED METHOD

The main goal in this work is to identify frame type of each video packet, without analyzing its payload. This can be done by inspection of certain characteristics of the video, which are explicitly explained in this section.

There are three main video frame types: I, P and B. Intra-frames or I-frames are coded without reference to other frames. P-frames are coded with predictions from previous frames, and B-frames are predicted bi-directionally. Each of these frame types generates different bits per frame. In general, they can be ordered in terms of bits/frame as: I-frame, P-frame and B-frame. If the size of each frame was known, an observer could easily tell the type of each passing frame in the network. But due to reasons such as error resiliency and MTU size limit, video frames are almost always sent in smaller parts (slices). Therefore, a network node cannot know the video packet type because it does not have the time and resources to query into its higher layers. Our method aims to enable the network to reach a frame-level understanding of the video only from the IP header data of the video packets.

It is assumed that each slice of a video frame is encapsulated into exactly one RTP packet which is transmitted over UDP/IP in a network that guarantees connection-oriented transmission. This scheme is most common for delay-sensitive live video streams [37]. Although the method is considered to work over UDP but under certain conditions, it can also work on TCP. Most TCP implementations are not suitable for live video because they enforce in-order delivery, retransmission and rate adjustments. But there are variants of TCP that are modified to suit multimedia applications, TCP Hollywood [38] is one such example. In general, as long as the transport protocol is video-friendly, the method is applicable.

Generally, there are two ways to slice a frame where each slice is transmitted as a packet. They are: either equal number of coding units (macroblocks) per slice, or equal number of bits/bytes per slice. In the former, based on frame type, length of packet payloads will differ (I: largest, B: smallest). In the latter, since slices have almost the same size, their inter-generation times, or equivalently their packet inter-arrival times into a network node will be different. I-slices, which are less compressed, are made up of fewer macroblocks. Thus, their inter-generation time would be the shortest. Conversely, B-slices are efficiently coded and their packets are made up of more macroblocks, making their inter-generation time longer.

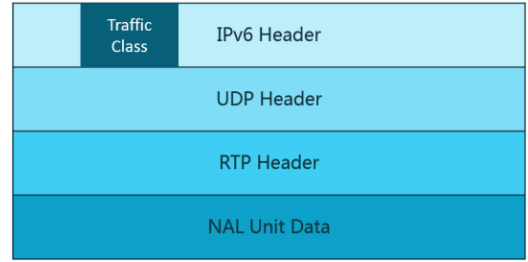

Fig. 1. An IPv6 packet structure for live video streaming

For each of these structures, a different algorithm is used to extract the frame features, which are described in parts A and B of this section. Then, in part C, it is explained how the classification is carried out.

To avoid the various effects of network routing protocols on the video sequence, the method can be implemented at the very first routing capable network node. After that, the packets header (e.g. traffic class) can be marked based on their importance and this information can be carried out to other nodes on the path to the destination point.

### A. Fixed Number of Coding Units per Slice

When each slice is comprised of fixed number of coding units or macro-blocks, frames have constant number of slices. Therefore, the first step for revealing the video structure is to find how many slices are in each frame, which results in finding the frames size.

In each video frame, moving forward by $N$ slices from any slice, when $N$ is equal to the number of slices per frame, we will reach the same position in the next frame. In adjacent frames, due to spatio-temporal correlation, slices in the same position are very similar in size if their types are the same. Since video content remains unchanged for some time [33] (i.e. that is how rate allocation is done in video codecs), slice sizes in consecutive frames change coherently, e.g. from an I-frame to a neighboring B-frame, all the slices undergo size reduction.

Consider two adjacent non-overlapping windows, $W_{i,n}$ and $W_{i+1,n}$ that are put over video frames and each contains $n$ packets (slices). The decision function $D_n$ is defined as:

$$D_n = \frac{(\mu_{i,n} - \mu_{i+1,n})^2}{\sigma_{i,n}^2 + \sigma_{i+1,n}^2} \quad (1)$$

where $\mu_{i,n}$ is the average and $\sigma^2_{i,n}$ is the variance of the bits in the $i^{th}$ window with $n$ slices.

At the beginning and when the window length is small ($i=1$, $n \ll N$), most likely both windows cover slices from the same frame, and hence their average, $\mu$, will be similar, leading to small $D_n$. By increasing the window length ($n$), $W_{1,n}$ will cover larger portion of the current frame and when it reaches its largest value, it forces $W_{2,n}$ to enter into the next frame. According to the explained logic, after repeating the process for all window sizes, $D_n$ is maximized when each window contains only the same type slices (minimum variance) and these types are different between the windows (maximum average difference). This indicates a frame type change and therefore $n$ is equal to $N$.

In order to verify the value of *n*, the operation is repeated for *i*=2,3,4,… until the same result is observed for at least 4 consecutive times. As this method finds the frames border, it does not matter if the first packet in $W_{1,n}$ is the first slice of the frame or not. After finding *N*, frames sizes can be calculated. (i.e. sum of slice sizes in that frame).

### B. Fixed Number of Bytes per Slice

In case of constant size slices (packets), each frame will consist of different random number of slices and as a result the number of packets in each frame will be different. Therefore, to find the video structure, the method explained in section A cannot be pursued.

Had the packets had exactly the same size, the last packet of each frame would have been smaller than the others and could be used as a frame change flag. However, in practice such packets are normally padded with extra bits that are ignored by the receiver. As a result, the last packet of frame cannot be recognized by its size. In this scheme, inter-arrival time of packets can be used to find the frame borders.

Due to content diversity among slices, packets inter-arrival times change stochastically even among the same type slices. This is more severe in I packets, which cover smaller portions of a scene and there is a high probability that a portion of I-frame will be completely plane or has complex texture. Moreover, since packet inter-arrival times in practice will be carried out at the router, then network queueing will impose a random jitter on the inter-arrival times. Of course, encoders may employ traffic smoothing buffer, to combat against these fluctuations in the packet sending rate. To avoid these anomalies and attain a more extensive model, the average of inter-arrival time values should be used, which highlights the importance of locating the last slice of frames even more.

Since I-frames are less compressed, coding complexity is low and filling a packet requires fewer building blocks, i.e. the inter-generation time of I-slices or the inter-arrival times of I packets are shorter than those of P packets. Similarly, B packets will have the largest inter-arrival times. As macroblocks are packed into a packet in whole numbers, then the packets size may slightly differ. Therefore, in order to better reflect different frame type characteristics, packet size is used alongside its inter-arrival time.

For this purpose, packet size to its inter-arrival time ratio is used, which is called bit-generation ratio and is defined as:

$$R_i = \frac{S_i}{T_i} \qquad (2)$$

where $S_i$ and $T_i$ are the size and inter-arrival time of the $i^{\text{th}}$ packet, respectively.

Consider a moving window $W_i$ of constant length $2L+1$, that is moved over the $R_i$ values from a random starting packet. A decision function is defined as:

$$C_i = \frac{(\lambda_i)^2 * (\mu_{i,1} - \mu_{i,2})^2}{\sigma^2_{i,1} + \sigma^2_{i,2}} \qquad (3)$$

where $\lambda_i$ is the size difference of the middle packet from the average size of the packets. $\mu_{i,1}$ and $\mu_{i,2}$ are the mean $R_i$ values of the left and right groups, respectively, and $\sigma^2_{i,1}$ and $\sigma^2_{i,2}$ are their variances. The middle element of window $W_i$ for which $C_i$ is greater than a predefined threshold would be a candidate for the last packet of a frame. Larger values of $C_i$ indicate that the central tendencies of the two groups are maximally distanced and their variances are minimum, which is equivalent to frame type change, just like the logic explained in section A.

At this stage, the average $R_i$ of all the packets between any two consecutive frame type change candidates will be the output of this section.

### C. Frame Type Identification

After feature extraction in sections A and B, K-means clustering is used for frame type classification. K-means is a simple and efficient clustering algorithm [39]. In K-means, the order of data is not important, therefore, it can operate fast and accept data in batches. In the proposed method, K-means categorizes the frames into different clusters and each cluster represents a frame type. To match the clusters with actual frame types, clusters average is used. The frames in the cluster with the largest/ average value are labeled as I-frame. The next largest value belongs to the P-frames, and the smallest value will be for the B-frames. An algorithmic procedure is summarized in **Algorithm 1**, where it describes how the method identifies the packet type based on the packets inter-arrival time.

In the algorithm, functions *avg(x, y, z)* and *var(x, y, z)* return the average and variance of *z* members of *x* vector starting from *y* respectively. Finally, the values in *packets_type* are mapped

---

**Algorithm 1**: Packet type identification when each slice is comprised of fixed number of bytes

**Input:** *N* packets with *Size* and Inter-arrival time *(IAT)* and a *THRESHOLD* value
**Output:** Obtained *packets_type*
1: **for each** *i* ∈ [0, N-1] **do**
2:     *R[i] = Size[i] / IAT[i]*
3: **end for**
4: *m = avg(Size, 0, N)*
5: **for each** *i* ∈ [0, N-1] **do**
6:     λ = 1
7:     **if** *Size[i] != m*
8:        λ = *m - Size[i]*
9:     **end if**
10:    *C[i] =* power2(λ) * power2(avg(*R, i+1, l*) – avg(*R, i-l, l*)) / (var(*R, i+1, l*) + var(*R, i-l, l*))
11: **end for**
12: *j = 0*
13: *b = 0*
14: **for each** *i* ∈ [0, N-1] **do**
15:    **if** *C[i] > THRESHOLD*
16:      *Frames_feature[j++]* = avg (*R, b, i-b*)
17:      *b = i*
18:    **end if**
19: **end for**
20: *clusters =* k-means (*Frames_feature*, 4)
21: *k = 0*
22: j=0
23: **for each** *i* ∈ [0, *N*-1] **do**
24:    *packets_type[i] = clusters[j]*
25:    **if** *C[i] > THRESHOLD*
26:      *j+=1*
27: **end for**
28: **return** *packets_type*

to I, P, and B frames.

It is noteworthy that since content and encoding parameters are seldom changed during a single stream, there is no need to run the algorithms continuously and the node can classify new packets based on observations from only a small portion of the video. Once the method finds the cluster averages, it labels the new packets based on their Euclidean distance from the clusters.

When the number of slices in video frames is constant, revealing the GoP (Group of Pictures) structure helps in labeling new packets without needing to run the algorithm. Video pattern is repeated in GoP structures and adjacent GoPs have similar characteristics: e.g. size. This will result in identifying to which frame type each packet belongs to. In fixed number of coding units per slice, as each GoP has constant number of slices, finding GoP size has two advantages. 1) It can increase classification accuracy by optimizing the input vector and 2) it enables the method to exploit packet orders (transmission or display) to find the frame types. Therefore, after finding the video GoP structure, running the explained algorithm will be unnecessary. As will be explained below, by knowing the frame size, it is feasible to reveal the GoP structure.

Consider a window of length $l$ which moves forward over the frames one-by-one and finishes when a total of $K$ frames are observed. As shown in equation 4, at each step, the absolute size of the last frame is subtracted from the first frame in the window, and is summed over the entire operation:

$$G_l = \sum_{i=1}^{K-l} |S_{i,1} - S_{i,l}| \quad (4)$$

where $S_{i,1}$ and $S_{i,l}$ are the size of the first and last frames in the $i^{th}$ step, respectively. Repeating this operation for various window lengths, ranging from $l=2$, to an arbitrary maximum value (e.g., $l_{max}=31$), when $l-1$ is equal to the actual GoP size, the first and last frames of the window become the same type and hence $G_l$ is minimized and GoP is found.

In the above-explained method variable $K$ is defined as the minimum number of frames required to find the correct window length. Value of $K$ depends on several parameters, such as: video content, GoP structure, GoP size, encoding parameters, etc. As the window must at least observe two GoPs, therefore $K >= 2l_{max}$.

For better visualization of the relation between GoP size and window length, Figure 2 shows a sequence of video frames with GoP size 12 in the display order. The window lengths in (a) and (b) are 13 and 12, respectively. When the window moves, in (a), the same frame types are subtracted from each other, leading to minimum value for $G_{13}$ but in (b), $G_{12}$ will be large because different frame types are selected.

In fixed number of bytes per slice method, since the number of packets in frames is not constant, the number of packets in each GoP is variable. Therefore, finding the GoP structure cannot help in future packet identification and only clusters averages are used.

*D. Hierarchical Coding Support*

In standard codecs, for better utilization of the temporal redundancy of the video, the number of B-frames per GoP can be increased. For efficient coding of large number of B-frames, some of them can be used as reference for the other B-frames. This structure is called hierarchical coding in which there are two B-frame types: reference (B) and non-reference (b), which are the least important frames. Discriminating between these two types can help maintain video quality better.

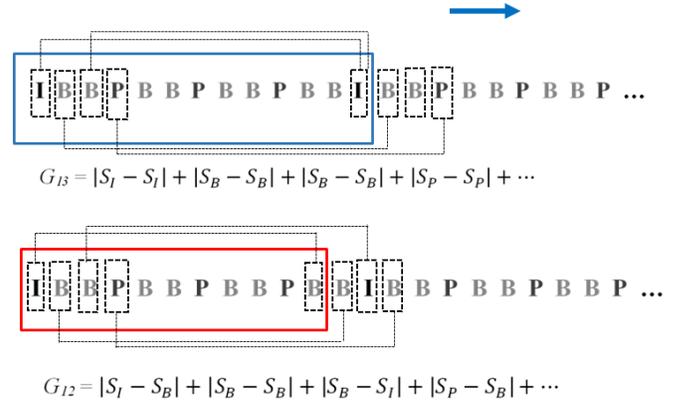

Fig. 2. For a GoP size of 12. In (a) Correct window length of 13, leads to identical frame type differences minimizing G, and (b) wrong window size of 12 leads to dis-similar frame type differences, increasing G.

Since characteristics of B and b packets are similar, clustering might not be efficient for separating them. In this case, information about packets order (transmission or display) can be used. For example, in the transmission/encoding order, reference frames are sent prior to the non-reference frames. Applying this rule is harmless because even if video structure is not hierarchical the last frames encoded in a GoP are still the least important packets.

*E. Complexity*

To measure the complexity of packet type identification program, the processing load of the workflow should be considered. At first, the program finds the video slicing method of the received bit-stream by comparing the packets size for a short period. In case of variable packet size, since the proposed method needs only $K$ frames to reveal the video structure and it is independent from the number of video frames, its computational complexity can be considered to be $O(1)$. Although, k-means has a complexity of $O(n^2)$, but it applies only to a few frames, and there is no need to run it for all the frames. For fixed size packets, calculating C and applying k-means algorithm has a complexity of $O(n^2)$. Therefore, the computational complexity of the program will be an order of $n^2$.

As for the storage complexity, since only one value is extracted from each packet and there is no need to keep the packets, the memory space is needed only as much as to keep a track of the studied packets, e.g. for the method explained in section A, $K \times N$ integer values should be kept in device memory.

## IV. EXPERIMENTAL RESULTS

Six test video sequences (i.e. Ducks Take Off, Four People, Johnny, Old Town, Park Joy, and Shields) in HD format (1280×720 pixels) were encoded by JM19.0 software of H.264 standard video codec at 30 frames/s and constant bit-rate of

TABLE I
GOP STRUCTURES USED IN THE SIMULATIONS (DISPLAY ORDER)

| Profile Number | N | M | GoP Structure |
|---|---|---|---|
| 1 | 12 | 3 | I B B P B B P B B P B B |
| 2 | 15 | 3 | I B B P B B P B B P B B P B B |
| 3 | 15 | 5 | I B B B B P B B B B P B B B B |
| 4 | 18 | 3 | I B B P B B P B B P B B P B B P B B |
| 5 | 18 | 6 | I b B b B b P b B b B b P b B b B b |

TABLE II
AVERAGE ACCURACY OF PACKET IDENTIFICATION ALGORITHM (FIXED NUMBER OF PACKETS PER FRAME)

| Actual \ Predicted | I | P | B | b |
|---|---|---|---|---|
| I | 100% | 0.0% | 0.0% | 0.0% |
| P | 0.0% | 99.1% | 0.9% | 0.0% |
| B | 0.0% | 0.2% | 99.8% | 0.0% |
| b | 0.0% | 0.1% | 0.0% | 99.9% |

TABLE III
AVERAGE ACCURACY OF PACKET IDENTIFICATION ALGORITHM (FIXED NUMBER OF BYTES PER PACKET)

| Actual \ Predicted | I | P | B | b |
|---|---|---|---|---|
| I | 98.3% | 1.7% | 0.0% | 0.0% |
| P | 3.1% | 90.2% | 3.7% | 3.0% |
| B | 0.0% | 12.1% | 79.9% | 8.0% |
| b | 0.0% | 3.0% | 19% | 78.0% |

4Mbit/s. Table 1 shows the GoP structure of the coded video sequences in display order. The value of *N* specifies their GoP size, and the value of *M* determines the distance between reference pictures. All the video sequences were encoded into either 100 slices per each frame (method 1) or 1024 bytes per each slice (method 2). Please note at the routers, packets are received in the transmission/ encoding order, and their GoP structure representation in this order for instance in profile number 5, will be in the order of: IPBBbbbP…PBBbbb.

The identification program was written in C++ programming language. Also, to simulate packet discard scenario, another code was developed that uses the identification program decisions and drops video packets based on the required bandwidth and packet frame type. The results are presented in two parts: identification accuracy and video quality preservation.

*A. Identification Accuracy*

To evaluate the accuracy of the identification methods, encoded videos are passed to the identification program and the outputs are compared with the actual packet types. Table 2 is the confusion matrix for fixed number of slices per frame, and hierarchical B-structure. The diagonal elements of the table show the accuracy of frame type detection. As can be seen, I-packets are never mistaken for any other frames. This is a favorable outcome, since the most-important packets of the stream are correctly detected and will never be lost in the network. For P and B-frames wrongly detected packets are less than 1%.

The packet type prediction accuracy in a more practical environment of network jitter at a constant size packets of 1KB with and without smoothing buffer are evaluated. Coded video at 4 Mbit/s and 1 KB packets, will have 500 packets per second. With a normal GoP size of 15 frames and the video frames complexity factor of 5, 3 and 1 for I, P and B-frames respectively, on average each I-frame has 46.29 packets, which means their average inter-arrival time will be 720 µs. Jitters are usually modelled as a truncated Poisson distribution, which its skewed shape looks more like a normal distribution. Hence for simplicity normal distributions are added to the packets inter-arrival times at three different parameters of: ($\mu_1$=72, $\sigma_1$=24), ($\mu_2$=144, $\sigma_2$=48), ($\mu_3$=288, $\sigma_3$=96). The identification accuracy when hierarchical coding is enabled in the codec configurations (profile number 5) and a random jitter with a mean 72 µs is added to packets inter-arrival time, is shown in Table 3.

In random jitter with the mean of 72 µs, the accuracy for I packets is very high and these packets are never mistaken with B packets.

Table 4 shows the confusion matrix for the explained scheme with normal B-frames when different jitter values are applied. As the jitters increase, the probability of mistaking an I packet for a P packet increases. This is because I packets have smaller inter-arrival times and the jitter has more effect on them. However, I packets are still rarely mistaken with B packets, which is very important because B packets are the first to be discard in case of congestion.

To test the effect of smoothing buffer, the above experiments are repeated, but this time a smoothing buffer is put at the encoder output. The buffer waits until all the packets of a frame are generated and then transmits them one by one so that the time interval between sending the first and the last packet of all the frames is equal. Table 5 shows the results of identification for this scheme. As can be seen, applying the smoothing buffer will ensure that the effect of increasing the jitter value does not lead to packet type misidentification.

In Table 5, moving down the main diagonal, the inter-arrival of packets become closer and therefore the accuracy decreases. However, wrong detections are concentrated under the main diagonal, which means that less important packets are mistaken for the more important ones. This only removes some B-frames from the discard list and important packets are still kept untouched, which preserves video quality.

What is harmful, is when an I or a P-frame is mis-predicted as a B-frame. This case is very rare and as shown in the tables, it happens less than 0.9%, 0.6% and 5.6% of the times for methods 1, 2 with and without smoothing buffer respectively.

The reason for the better performance of method 1 over method 2 when smoothing buffer is not used, is that in highly textured areas, some macroblocks in P- and B-frames, might be intra coded, reducing their inter-arrival times, and in plain areas, blocks in P-frames may not generate enough bits, prolonging their inter-arrival times, making them to look like B-frames.

TABLE IV
AVERAGE ACCURACY OF PACKET IDENTIFICATION ALGORITHM WITH DIFFERENT JITTER VALUES

| The jitter mean value | 72 µs | | | 144 µs | | | 288 µs | | |
|---|---|---|---|---|---|---|---|---|---|
| | I | P | B | I | P | B | I | P | B |
| I | 98.2% | 1.8% | 0.0% | 89.8% | 10.2% | 0.0% | 87.3% | 12.6% | 0.1% |
| P | 1.2% | 95.4% | 3.4% | 0.4% | 95.8% | 3.8% | 2.6% | 91.9% | 5.5% |
| B | 0.0% | 7.9% | 92.1% | 0.0% | 7.9% | 92.1% | 0.0% | 8.8% | 91.2% |

TABLE V
AVERAGE ACCURACY OF PACKET IDENTIFICATION ALGORITHM WITH DIFFERENT WHEN SMOOTHING BUFFER IS APPLIED

| The jitter mean value | 72 µs | | | 144 µs | | | 288 µs | | |
|---|---|---|---|---|---|---|---|---|---|
| | I | P | B | I | P | B | I | P | B |
| I | 99.7% | 0.3% | 0.0% | 99.6% | 0.4% | 0.0% | 96% | 4.0% | 0.0% |
| P | 0.5% | 98.9% | 0.6% | 1.2% | 98.3% | 0.5% | 1.8% | 98.0% | 0.2% |
| B | 0.0% | 3.6% | 96.4% | 0.0% | 3.3% | 96.7% | 0.0% | 4.9% | 95.1% |

TABLE VI
COMPARISON OF PSNR RESULTS BETWEEN THE RANDOM, GRAND TRUTH AND THE SMART METHOD FOR DIFFERENT DROP PROFILES.

| Video | M, N | Intact PSNR (dB) | Low | | | Middle | | | High | | |
|---|---|---|---|---|---|---|---|---|---|---|---|
| | | | Grand Truth | Random | Proposed | Grand Truth | Random | Proposed | Grand Truth | Random | Proposed |
| Ducks_take_off | (12,3) | 27.74 | 27.5 | 25.03 | 27.38 | 27.18 | 21.87 | 26.72 | 27.49 | 16.92 | 25.66 |
| | (15,3) | 27.79 | 27.54 | 25.14 | 27.38 | 27.16 | 22.03 | 26.69 | 27.54 | 17.04 | 25.25 |
| | (15,5) | 27.36 | 27.1 | 24.74 | 27.02 | 26.61 | 21.78 | 26.5 | 25.32 | 17.17 | 24.94 |
| | (18,3) | 27.82 | 27.56 | 25.1 | 27.36 | 27.13 | 22.27 | 26.55 | 27.78 | 17.21 | 24.61 |
| | (18,6) | 28.08 | 27.54 | 25.58 | 27.06 | 26.39 | 22.31 | 25.66 | 23.87 | 17.65 | 23.19 |
| Four_prople | (12,3) | 42.83 | 42.35 | 33.05 | 41.4 | 41.49 | 27.86 | 40.12 | 39.46 | 22.57 | 37.15 |
| | (15,3) | 43.01 | 42.53 | 32.8 | 42.04 | 41.67 | 27.97 | 40.59 | 39.67 | 22.54 | 37.29 |
| | (15,5) | 42.63 | 42.09 | 33.18 | 41.33 | 41.16 | 27.22 | 40.03 | 38.76 | 21.57 | 37.92 |
| | (18,3) | 43.14 | 42.64 | 33.7 | 41.95 | 41.82 | 28.71 | 40.53 | 39.73 | 22.53 | 36.79 |
| | (18,6) | 43.25 | 42.86 | 31.37 | 42.77 | 42.22 | 25.63 | 41.97 | 40.57 | 19.83 | 39.15 |
| Johnny | (12,3) | 43.37 | 43.12 | 33.24 | 42.08 | 42.71 | 28.5 | 40.89 | 41.61 | 23.08 | 38.11 |
| | (15,3) | 43.52 | 43.28 | 33.12 | 42.82 | 42.87 | 29.49 | 41.83 | 41.76 | 23.36 | 38.55 |
| | (15,5) | 43.15 | 42.85 | 33.04 | 42.37 | 42.34 | 27.67 | 41.29 | 40.98 | 22.23 | 36.55 |
| | (18,3) | 43.63 | 43.39 | 35.32 | 42.26 | 42.97 | 28.75 | 41.12 | 41.86 | 23.6 | 37.98 |
| | (18,6) | 43.54 | 43.41 | 30.3 | 43.36 | 43.17 | 25.08 | 43.06 | 42.49 | 19.15 | 41.94 |
| Old_town | (12,3) | 37.78 | 37.03 | 27.82 | 35.36 | 35.81 | 24.39 | 34.3 | 32.99 | 20.48 | 29.93 |
| | (15,3) | 37.93 | 37.13 | 27.55 | 34.79 | 35.89 | 24.68 | 34.45 | 33.08 | 20.42 | 29.87 |
| | (15,5) | 37.59 | 36.86 | 28.5 | 35.42 | 35.63 | 23.56 | 33.68 | 32.72 | 20.03 | 29.31 |
| | (18,3) | 38.04 | 37.23 | 28.38 | 35.88 | 35.97 | 24.23 | 32.76 | 33.13 | 20.86 | 28.48 |
| | (18,6) | 38.03 | 37.27 | 29.99 | 37.12 | 36.08 | 25.32 | 34.73 | 35.11 | 19.82 | 29.47 |
| Park_joy | (12,3) | 27.44 | 26.57 | 24.67 | 26.21 | 25.2 | 21.68 | 24.57 | 24.28 | 17.54 | 21.88 |
| | (15,3) | 27.55 | 26.72 | 24.78 | 26.14 | 25.35 | 21.73 | 24.18 | 24.38 | 17.47 | 21.37 |
| | (15,5) | 27.3 | 26.66 | 24.75 | 26.44 | 25.55 | 21.82 | 25.32 | 22.9 | 17.57 | 22.17 |
| | (18,3) | 27.63 | 26.79 | 24.68 | 26.65 | 25.46 | 21.61 | 24.96 | 23.9 | 17.45 | 23.22 |
| | (18,6) | 27.71 | 26.92 | 24.98 | 26.86 | 25.69 | 21.92 | 25.42 | 23.37 | 17.81 | 22.08 |
| Shields | (12,3) | 35.56 | 34.28 | 23.88 | 32.04 | 32.32 | 20.11 | 27.51 | 28.7 | 16.74 | 23.94 |
| | (15,3) | 35.74 | 34.44 | 23.99 | 32.18 | 32.46 | 20.34 | 28.39 | 28.77 | 16.8 | 23.24 |
| | (15,5) | 35.45 | 34.27 | 24.89 | 32.9 | 32.37 | 21 | 31.25 | 28.42 | 17.11 | 25.78 |
| | (18,3) | 35.83 | 34.52 | 24.11 | 32.52 | 32.53 | 20.25 | 27.84 | 28.86 | 16.81 | 23.38 |
| | (18,6) | 35.93 | 34.53 | 27.81 | 34.49 | 32.5 | 23.31 | 32.41 | 25.85 | 18.83 | 24.44 |

In order to evaluate the performance of smart drop method in preserving the video quality three discarding scenarios are considered, which are thoroughly discussed in the next section.

### B. Packet Discard Scenario

Although method 1 gives more accurate frame-type detection than method 2, but in practice fixed number of bytes/slice (method 2) is more likely to be used. Therefore, for discarding scenarios only method 2 is tested. Also, for the packets size, the value of 1024 bytes was chosen. This is a reasonable value because it is not too small to increase the overhead of the video packets and it is smaller than the common network MTU limit. Also, a random jitter value is added to the inter-arrival time of each packet and as the worst case scenario the smoothing buffer was not used in these experiments.

It is assumed that the video sequences enter a hypothetical network node one at a time and the node is over-loaded to three different extents (low, medium, high) such that respectively 1%, 3% and 10% of the bit-rate of each video stream should be dropped. For each packet loss rate, three different approaches were tested: (i) random packet drop; (ii) packet drop based on absolute knowledge of the packet type (ground truth) and (iii) packet drop based on the proposed method. In the proposed smart drop and ground truth drop, packets are dropped in the priority order of b, B, P and I-frame types. Like the experiments in previous part of this section, a random jitter with the mean of 72 µs and σ=24 (±3σ corresponds to 99.9% confidence limit) was added to packets inter-arrival times.

The reconstructed video quality was measured in terms of PSNR. As claimed in [40], PSNR is a valid quality measure to compare videos at various quality levels, as long as its content remains the same. This is supported by a recent investigation, where the performance of 13 variety of image/video quality metrics, including PSNR, variants of SSIM, VQM.., etc. were extensively studied for measuring the perceived visual quality of loss concealed video, and surprisingly for a set of standard test video sequences, PSNR has been either the closest or the second closest method to the subjective tests mean opinion scores [41].

Table 6 lists the PSNR of the decoded video sequences under various GoP lengths of Table 1, for the ground truth, random drop and the proposed method. The absolute PSNR value of each video is shown in the second column (intact). Detail inspection of this Table highlights the following points:

- The proposed method closely follows the ground truth performance. For Instance, in the low scenario the average difference between the PSNR of both methods is 0.78 dB. This value for the medium and high scenario is 1.31 dB and 2.69 dB respectively.
- In high drop rates, the performance difference may rise to up to 4-5 dB. Because all non-reference packets that the proposed method did identify are dropped and it has to drop from reference frames, which it causes drift distortion and quality loss. Nevertheless, in these cases, the proposed method is still 7 dB better than the random method and in some cases, it outperforms random drop method by more than 20 dB.

## V. CONCLUSION

In this paper a method for video packet type identification in network was introduced. The goal was to identify I, P and B-frame type packets based on the information extracted from the packet header. This information comes from inherent characteristics of the encoded video which are reflected in either the packet size or the inter-arrival time based on the packetization scheme.

In the first scheme, video frames are partitioned into equal size slices and hence the generated video packets size is varied. In the second scheme, compressed video slices are packetized into constant size packets and hence their inter-arrival time is varied. It was shown that in both schemes reference slices are identified with 95.1% accuracy. However, in practice the second scheme of equal size packets is more commonly used.

Through a variety of test video sequences and GoP structures, it was shown that the performance of the proposed packet drop system follows that of the ground truth closely. This makes sure the identified packets are as accurate as if they were truly tagged by an honest sending host.

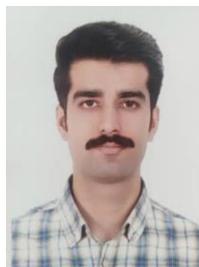

**Ashkan Moharrami** recieved the B.Sc Degree in electrical engineering from the International University of Imam Khomeini, Qazvin, Iran in 2015 and the M.Sc in electrical engineering from the University of Tehran, Tehran, Iran in 2018. As of now he is a researcher at the Multimedia Processing Laboratory of the University of Tehran. His main research interests are: Video over Data and Cellular Networks.

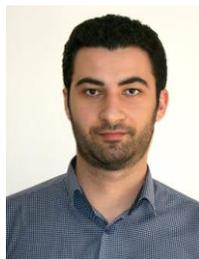

**Mohammad Ghasempour** graduated from University of Semnan, Iran in 2016 and his Master degree in Computer architecture from University of Tehran, Iran in 2018. He is currently a Ph.D. candidate at the School of Computer and Electronic Engineering University of Tehran, Iran. His main research interests are: Video Processing and Computer architectures in video coding.

**Mohammad Ghanbari** (M'78, SM'97, F'01, LF14) is a professor of video networking at the University of Tehran, Iran and Emeritus Professor at the University of Essex, United Kingdom. He is best known for his pioneering work on two-layer video coding (which earned him an IEEE Fellowship in 2001), He is the author of seven books and his book: An Introduction to Standard Codecs, received the year 2000 best book award by IET. Prof. Ghanbari has authored or co-authored more than 750 journal and conference papers, thirteen patents, many of which have had a fundamental influence in the field of video networking. He was one of the founding Associate Editors to IEEE Transactions on Multimedia (IEEE-T-MM) and had served from 1998-2002. On January 2014 he was honored to Life Fellow of IEEE.

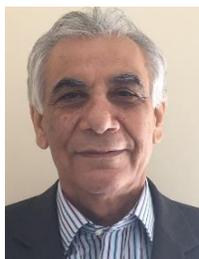